\documentclass[preprint,5p,numbers,sort&compress]{elsarticle}

\usepackage{graphicx}
\usepackage{dcolumn}
\usepackage{bm}
\usepackage{longtable}
\usepackage{float}
\usepackage{threeparttable}
\usepackage[colorlinks = true, citecolor=blue,urlcolor=blue,linkcolor=blue]{hyperref}
\usepackage{color} 
\usepackage{booktabs}

\usepackage{supertabular}%
\usepackage{float}

\usepackage{url}
\usepackage{array}
\usepackage{amssymb}
\usepackage{dblfloatfix}
\usepackage{tabu}
\usepackage{amsmath}
\usepackage{mathtools, cuted}
\usepackage[english]{babel}
\usepackage{lipsum}

\begin{document}


\title{Long-sought isomer turns out to be the ground state of $^{76}$Cu}

\author[jyv,sur]{L. Canete\corref{cor1}}
\ead{laetitia.canete@nulondon.ac.uk}
\cortext[cor1]{Corresponding author at: Northeastern University London, Devon House, 58 St Katharine's Way, E1W 1LP, London, United Kingdom}

\author[gan,nsc]{S. Giraud}


\author[jyv]{A. Kankainen\corref{cor2}}
\ead{anu.kankainen@jyu.fi}
\cortext[cor2]{Corresponding author at: University of Jyv\"{a}skyl\"{a}, P.O. Box 35, FI-40014 University of Jyv\"{a}skyl\"{a}, Finland}

\author[gan]{B. Bastin}

\author[str]{F. Nowacki}

\author[bor]{P. Ascher} 

\author[jyv]{T. Eronen} 

\author[gan,sac]{V. Girard Alcindor} 

\author[jyv]{A. Jokinen} 

\author[jyv,alt,hel]{A. Khanam}

\author[jyv]{I.D. Moore} 

\author[jyv]{D. Nesterenko}

\author[gan]{F. De Oliveira}

\author[jyv]{H. Penttil\"a}

\author[rom]{C. Petrone}

\author[jyv]{I. Pohjalainen}

\author[jyv]{A. De Roubin}

\author[jyv]{V.~Rubchenya\fnref{fn1}}

\author[jyv]{M.~Vilen}

\author[jyv]{J.~\"Ayst\"o}

\address[jyv]{University of Jyv\"{a}skyl\"{a}, Department of Physics, Accelerator Laboratory, P.O. Box 35(YFL) FI-40014 University of Jyv\"{a}skyl\"{a}, Finland}

\address[sur]{Department of Physics, University of Surrey, Guildford, GU2 7X5, United Kingdom}

\address[gan]{GANIL, Bd Henri Becquerel, BP 55027, F-14076 Caen Cedex 5, France}

\address[nsc]{Department of Physics and Astronomy, and the Facility for Rare Isotope Beams, Michigan State University,East Lansing, Michigan 48824-1321, USA}

\address[str]{Universit\'e de Strasbourg, CNRS, IPHC UMR 7178, F-67000 Strasbourg, France}

\address[bor]{LP2i, CNRS/IN2P3—Universit\'e, Bordeaux 1, 33175 Gradignan Cedex, France}

\address[sac]{Université Paris-Saclay, CNRS/IN2P3, IJCLab, 91405 Orsay, France}


\address[alt]{Department of Applied Physics, Aalto University, P.O. Box 11000, FI-00076 Aalto, Finland}

\address[hel]{Department of Physics, University of Helsinki, P.O. Box 43, FI-00014 Helsinki, Finland}

\address[rom]{IFIN-HH, P.O. Box MG-6, 077125 Bucharest-Magurele, Romania}

\address[cer]{Experimental Physics Department, CERN, CH-1211 Geneva 23, Switzerland}

\fntext[fn1]{Deceased}

\date{\today}

\begin{abstract}
Isomers close to the doubly magic nucleus $^{78}$Ni ($Z=28$, $N=50$) provide essential information on the shell evolution and shape coexistence far from stability. The existence of a long-lived isomeric state in $^{76}$Cu has been debated for a long time. We have performed high-precision mass measurements of $^{76}$Cu with the JYFLTRAP double Penning trap mass spectrometer at the Ion Guide Isotope Separator On-Line facility and confirm the existence of such a isomeric state with an excitation energy $E_x=64.8(25)$~keV. Based on the ratio of detected ground- and isomeric-state ions as a function of time, we show that the isomer is the shorter-living state previously considered as the ground state of $^{76}$Cu. The result can potentially change the conclusions made in previous works related to the spin-parity and charge radius of the $^{76}$Cu ground state. Additionally, the new $^{76}$Cu$(n,\gamma)$ reaction $Q$-value has an impact on the astrophysical rapid neutron-capture process.
\end{abstract}

\maketitle



\section{Introduction}
\label{sec:intro}
Long-living excited states known as isomers were discovered around 100 years ago \cite{Walker2020}. Today around 3340 nuclides and 1938 isomers with half-lives $T_{1/2}\geq$100~ns are known \cite{Kondev2021}. Isomerism is common in the region near doubly magic $^{78}$Ni, in the vicinity of the closed proton $Z=28$ and neutron $N=50$ shells. Although the magicity of $^{78}$Ni has recently been confirmed by experiments \cite{Olivier2017,Taniuchi2019}, the region still has many intriguing features, such as shape coexistence predicted via large-scale shell-model calculations \cite{Nowacki2016} and verified by experiments (see e.g. \cite{Yang2016,Nies2023}). 

Isomers can provide information on the shape coexistence phenomenon close to $^{78}$Ni. Laser spectroscopy on neutron-rich zinc ($Z=30$) isotopes with an odd mass number $A$ \cite{Wraith2017,Yang2016} have revealed an exceptionally large deformation in the $1/2^+$ isomer in $^{79}$Zn \cite{Yang2016} and indications of triaxiality in the $5/2^+$ isomer in $^{73}$Zn \cite{Wraith2017}. Isomeric states in the cobalt ($Z=27$) isotopes have been shown to have prolate-deformed intruder configurations coexisting with the normal near spherical states at low excitation energies \cite{Canete2020,Morales2017}. Spectroscopic studies on the odd-$A$ copper ($Z=29$) isotopes have shown that the position of the $\pi 1f_{5/2}$ orbital is sharply lowered as the occupancy of the $\nu 1g_{9/2}$ state increases. As a result, the single-particle $5/2^-$ state becomes the ground state at $^{75}$Cu ($N=46$) \cite{Franchoo1998,Flanagan2009,Olivier2017}. The collective ($1/2^-$) and ($3/2^-$) levels become very low in energy and produce nearly degenerate isomeric states in $^{75}$Cu  \cite{Daugas2010,Petrone2016,Ichikawa2019}. 

Isomerism is common in odd-odd copper isotopes but the evolution of the states has remained somewhat unclear as the shell-model predictions and experimental observations have disagreed, see e.g. Refs.~\cite{VanRoosbroeck2004,VanRoosbroeck2005,Vingerhoets2010}. The even-A $^{64,66,68,72}$Cu isotopes have a $6^-$ isomer which is the ground state in $^{70}$Cu  \cite{Vingerhoets2010,deGroote2020,Kondev2021,Singh2021}. The $6^-$ state is a member of the $(3,4,5,6)^-$ multiplet stemming from a $(\pi 1p_{3/2}\otimes \nu 0g_{9/2})$ coupling, {\it i.e.} from a neutron excitation across the $N=40$ subshell closure. $^{70}$Cu has two long-living isomeric states with spin-parities of $3^-$ and $1^+$ \cite{VanRoosbroeck2004prl,Vingerhoets2010}. The $1^+$ state has a dominant $(\pi 1p_{3/2}\otimes \nu 1p_{1/2})$ configuration, similar to the ground states of $^{64,66,68}$Cu \cite{Vingerhoets2010}. In $^{72,74}$Cu, the ground state has a spin-parity of $2^-$, proposed to originate from a $(\pi 0f_{5/2}\otimes \nu 1p_{1/2})$ coupling \cite{Vingerhoets2010,deGroote2020}.

For $^{76}$Cu ($N=47$), the existence of an isomeric state has been debated for a long time. So far, the only experiment that has reported two long-living states in $^{76}$Cu is a $\beta$-decay study of $^{76}$Cu at the TRISTAN online mass separator, where thermal neutrons on $^{235}$U and a plasma ion source were employed for the production \cite{Winger1990}. Winger $\it{et~al.}$ postulated that there has to be a high-spin ($J\approx 3-5$) and a low-spin ($J\approx 1-3$) state in $^{76}$Cu, with half-lives of 0.57(6)~s and 1.27(30)~s, respectively \cite{Winger1990}. The conclusion was based on the observed difference of half-lives for the 698-keV $4_1^+\rightarrow 2_1^+$ and 598-keV $2_1^+ \rightarrow 0^+$ beta-delayed $\gamma$-transitions in $^{76}$Zn. 

\begin {table*}
\caption{\label{tab:results} The frequency ratios $r=\nu_{c,ref}/\nu_c$, mass-excess values $ME_{JYFL}$ determined in this work via ToF-ICR measurements and half-lives $T_{1/2}$ from the literature for $^{76}$Cu. The spin-parities $I^{\pi}$ from \cite{Singh1995,Kondev2021} and the ground-state mass-excess value from \cite{AME20}, $ME_{AME20}$, are also listed. The excitation energy $E_x$ of the isomer was determined for the first time.}
\renewcommand{\arraystretch}{1.2}
\begin{tabular}{cccccccc}
\\
\hline
Nuclide  & $J^\pi$ & $T_{1/2}$ & $r$ & $ME_{JYFL}$ & $ME_{AME20}$ & Diff. & $E_x$  \\
 &  &  &  &  (keV) &  (keV) & (keV) &  (keV) \\
\hline

$^{76}$Cu 		& (1,2) \cite{Singh1995} & 1.27(30) s \cite{Kondev2021} & $0.905062917(26)$ & $-51011.4(20)$ & $-50981.6(9)$ & $-29.8(22)$ & -\\
$^{76}$Cu$^m$ & $3^-$ \cite{Kondev2021} & 637.7(55) ms\cite{Kondev2021} & $0.905063746(18)$ & $-50946.6(14)$ & - & - & 64.8(25) \\
\hline
\end{tabular}
\end{table*}

The existence of the 1.27(30)~s state in $^{76}$Cu was later disputed. Since the experiment of Winger $\it{et~al.}$ \cite{Winger1990}, all further experiments have failed to demonstrate the presence of another long-living state in $^{76}$Cu. Two $\beta$-decay studies of $^{76}$Cu, one performed at the ISOLDE facility at CERN \cite{VanRoosbroeck2005}, and one recently realized at the Holifield Radioactive Beam Facility (HRIBF) at Oak Ridge National Laboratory \cite{Silwal2022} observed no difference between the half-lives of the 599- and 698-keV transitions. The obtained half-lives, $T_{1/2}=653(24)$~ms \cite{VanRoosbroeck2005} and $T_{1/2}=637(20)$~ms \cite{Silwal2022}, agree with the high-spin, 0.57(6)~s state reported in Ref.~\cite{Winger1990}. Both experiments employed protons impinging into a uranium carbide target to produce $^{76}$Cu, combined with a laser \cite{VanRoosbroeck2005} or hot-plasma ion source \cite{Silwal2022}. Experiments performed using fragmentation reactions at the National Superconducting Cyclotron Laboratory \cite{Hosmer2010,Chester2021} also report only the shorter-living state in the $\beta$-decay of $^{76}$Cu. In addition, $\beta$-delayed neutron measurements at ISOLDE yielded a similar half-life of 641(6)~ms \cite{Kratz1991}. 

Furthermore, Penning-trap mass measurements using the Time-of-Flight Ion Cyclotron Resonance (ToF-ICR) technique \cite{Graff1980,Konig1995} at ISOLTRAP have not observed any indication of the possible existence of an isomeric state in $^{76}$Cu \cite{Guenaut2007,Welker2017}. The recent Collinear Resonance Ionization Spectroscopy (CRIS) study at ISOLDE reported a single state for $^{76}$Cu with a spin-parity $I^{\pi}=3^-$ \cite{deGrootePhD,deGroote2017}. The state observed in all experiments has been considered the ground state of $^{76}$Cu and the existence of the 1.27(30)-s isomeric state has been questioned \cite{Kondev2021}. In this Letter, we report on high-precision mass measurements on the ground and isomeric state in $^{76}$Cu and show that it is the ground state of $^{76}$Cu that has been missed in the previous experiments.

\section{Experimental method}
\label{sec:exp}
 
The $^{76}$Cu$^+$ ions were produced at the Ion Guide Isotope Separator On-Line (IGISOL) facility \cite{Moore2013} via fission induced by 35-MeV protons impinging into a 15 mg/cm$^2$-thick $^{nat}$U target. The fission fragments were stopped and thermalized in helium gas ($p\approx 310$~mbar), where the majority of the products ended up as singly charged ions. The ions were extracted out from the gas cell, guided through a sextupole ion guide (SPIG) \cite{Karvonen2008}, accelerated to 30 keV and mass-separated based on their mass-to-charge ratio $m/q$ with a 55$^\circ$ dipole magnet. The continuous beam was slowed down in a radiofrequency quadrupole (RFQ) cooler and buncher \cite{Nieminen2001}, which released the ions as bunches to the double Penning trap JYFLTRAP \cite{Eronen2012}. The $^{76}$Cu$^+$ ions were selected via the mass-selective buffer gas cooling technique \cite{Savard1991} in the first trap and transferred into the second trap, where the high-precision mass measurements took place using the ToF-ICR technique \cite{Graff1980,Konig1995}. 

\begin{figure}
\centering
\includegraphics[width=0.48\textwidth]{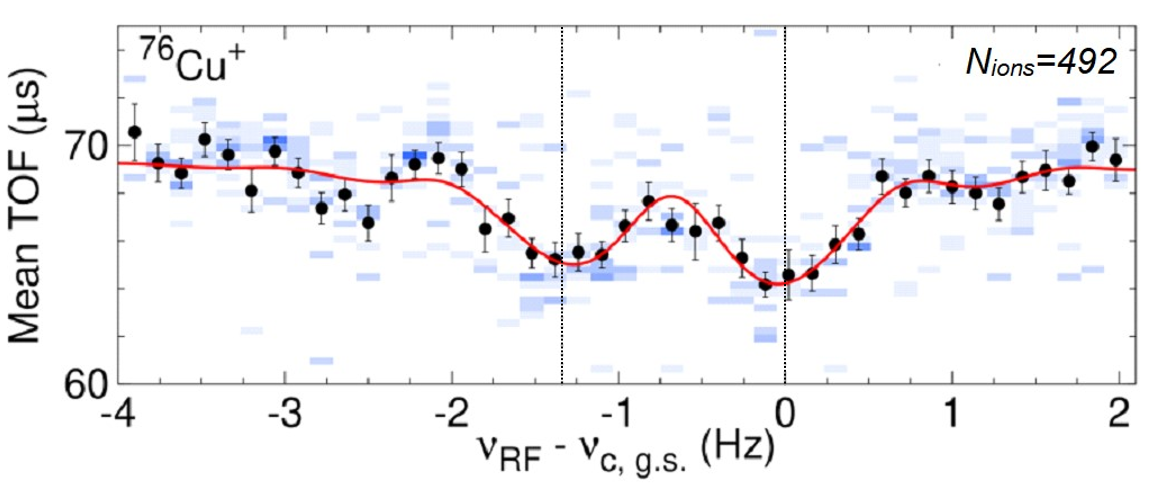} \caption{(Color online) Example of a ToF-ICR spectrum for $^{76}\mathrm{Cu}^+$ ($T_{RF}$=1120 ms). The background shading indicates the total number of ions, with darker shading meaning more ions. The solid red line is a fit of the theoretical curve on the data points. The vertical dashed lines show the resonance frequencies $\nu_c$ of the ground state (g.s.) and the isomer.}
\label{fig:toficr}
\end{figure}

\begin{figure}
\centering
\includegraphics[width=0.35\textwidth]{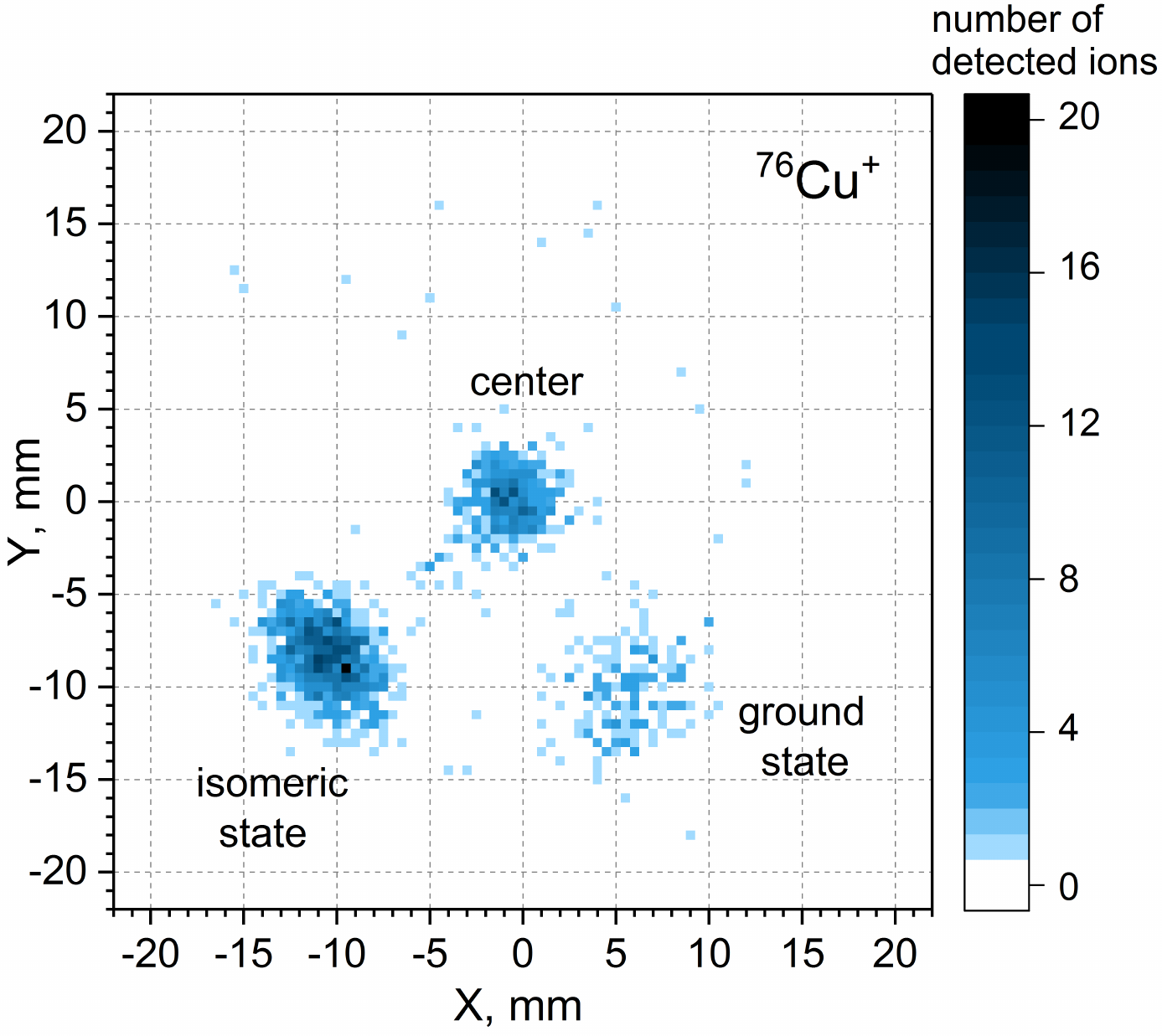} 
\caption{(Color online) The trap center and accumulated phase spots of the ground and isomeric states of $^{76}\mathrm{Cu}^+$ on the position-sensitive ion detector after 200~ms of phase accumulation time. The isomeric state is more dominantly produced in the fission reaction.}
\label{fig:pi-icr}
\end{figure}

A long quadrupolar excitation time of 1120 ms was used to achieve the high resolving power required to discriminate between the two close-lying states in $^{76}$Cu (see Fig.~\ref{fig:toficr}). The masses of the ground state and the isomer were derived from the measured cyclotron resonance frequencies, $\nu_c=qB/(2\pi m_{ion})$, where $m_{ion}$ is the mass and $q$ the charge of the ion and $B$ is the magnetic field strength determined using $^{84}$Kr$^+$ ions as a reference. Finally, the ground- and isomeric-state masses of $^{76}$Cu were determined as $m=r(m_{ref}-m_e)+m_e$, where $r=\nu_{c,ref}/\nu_c$ is the frequency ratio between the reference and the ions of interest, $m_{ref}$ the atomic mass of the reference  ($m(^{84}$Kr)= $83.911497727(4)~u$ \cite{AME20}) and $m_e$ the electron mass. Systematic uncertainties related to temporal fluctuations in the magnetic field strength, $\sigma_B(\nu_{c,ref})/\nu_{c,ref}$=$(8.18(19)\times 10^{-12}$/min$)\times\Delta t$~\cite{Canete2016}, and the mass-dependent uncertainty $\sigma_m(r)/r$=$(2.2(6)\times 10^{-10}/u)\times \Delta m$ \cite{Canete2019}, were quadratically added to the statistical uncertainties. Possible molecular contaminants and doubly-charged ions with $A/q=76$ were excluded as no matches were found for the determined frequencies.

\begin{figure}
\centering
\begin{tabular}{l}
\includegraphics[width=0.5\textwidth]{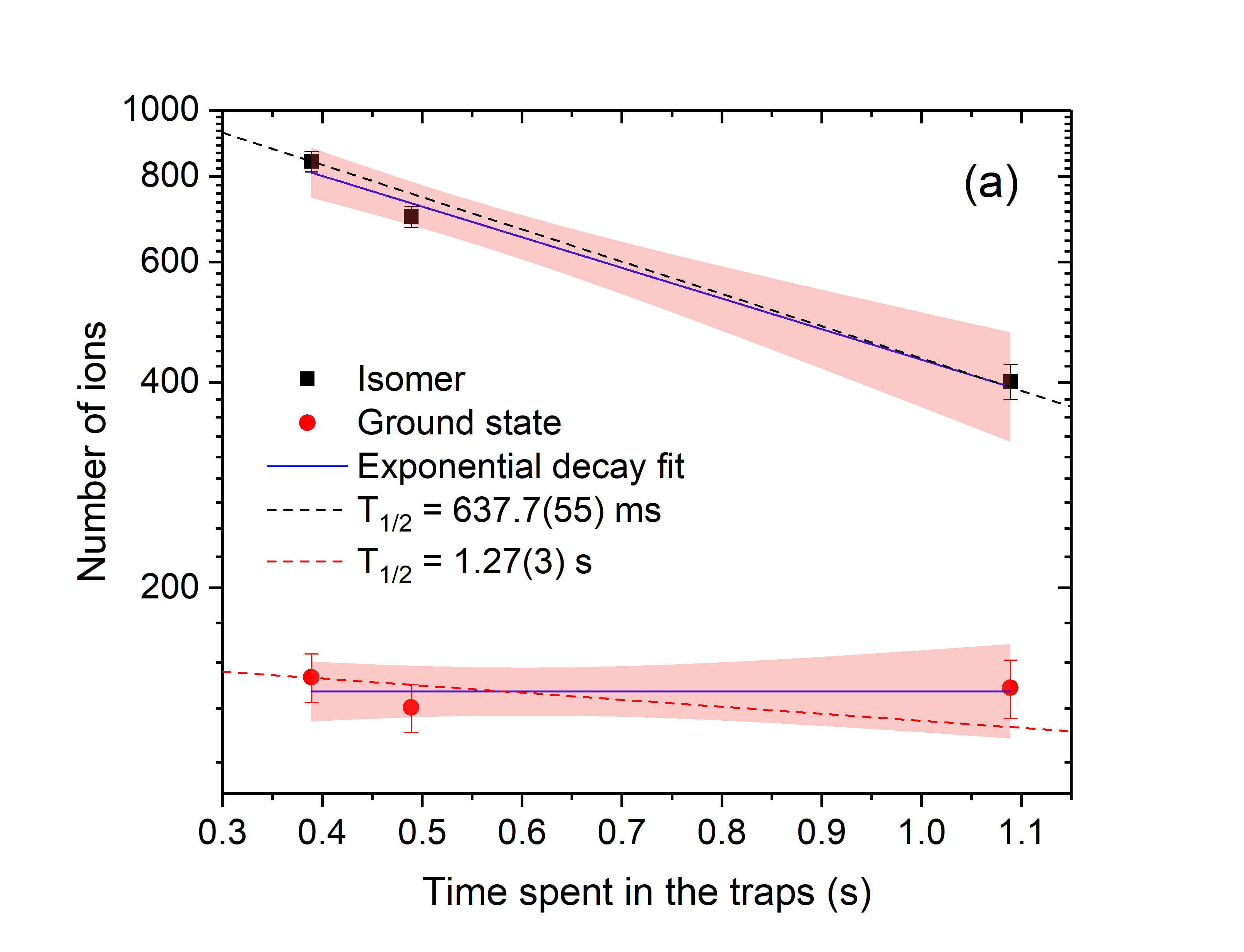} \\
\includegraphics[width=0.5\textwidth]{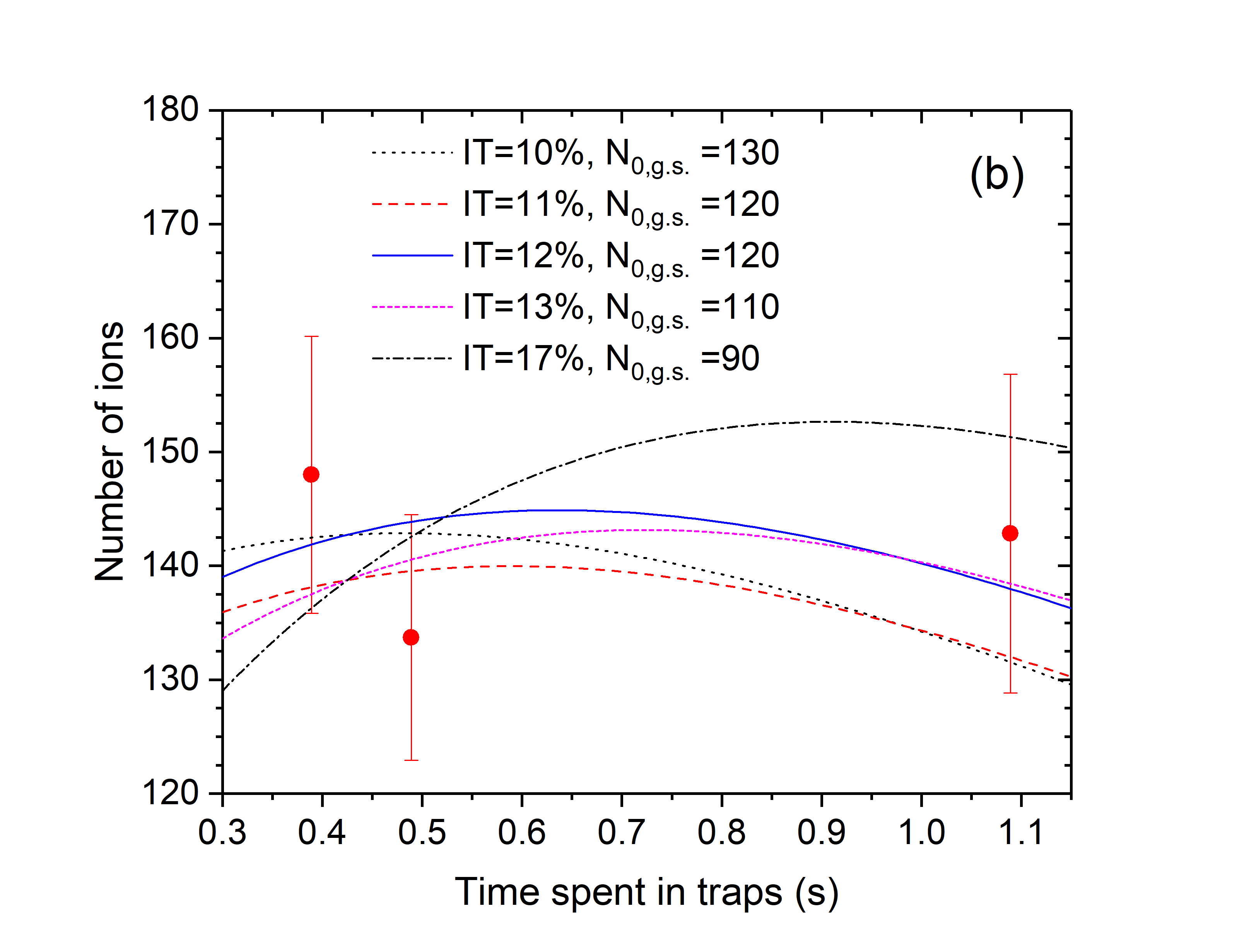}
\end{tabular}
\caption{(Color online) (a) The number of ions obtained via PI-ICR measurement decreases as a function of time spent in the trap according to the radioactive decay law. The fitted half-life (blue solid line) on the isomeric-state ions (black squares), $T_{1/2}$=672(110)~ms, is in agreement with the literature value, $T_{1/2}=637.7(55)$~ms \cite{Kondev2021} (black dashed line). The exponential decay fit on the ground-state ions (red circles) results in a flat trend, agreeing with the result $T_{1/2}=1.27(3)$~s from \cite{Winger1990} (dashed red line) only for the first two data points. Confidence bands (68~\%) of the fits have also been plotted. (b) The number of ground-state ions fitted using the literature half-lives \cite{Kondev2021,Winger1990} but different initial number of ions in the ground state ($N_{0,g.s.}$) and internal transition branch ($IT$) for the isomer. The data points are compatible with a dominant isomeric-state production $\gtrsim 90$~\% and an IT branch of around $10-17$~\% for the isomer.}
\label{fig:halflife}
\end{figure}

\begin{figure*}[h]
\centering
\includegraphics[width=1\textwidth]
{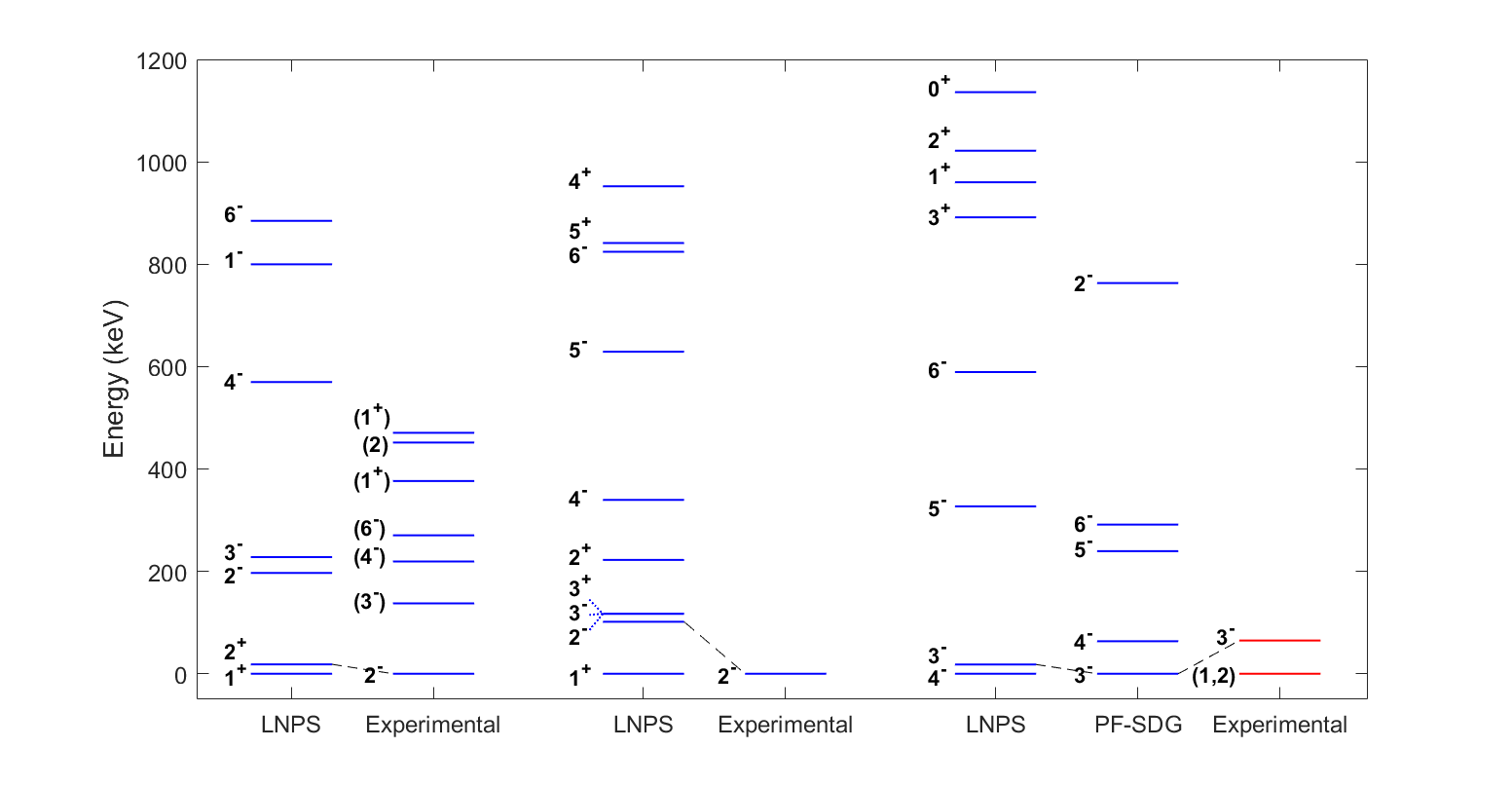} \caption{Experimental level schemes for $^{72}$Cu, $^{74}$Cu and $^{76}$Cu in comparison with shell-model calculations using the LNPS \cite{Lenzi2010} and PFSDG-U \cite{Nowacki2016} interactions. The energy of the levels in red have been measured in this work. The spin-parity assignment is tentative for the ground state of $^{76}$Cu. The experimental level energies in $^{72}$Cu are from Refs.~\cite{Flanagan2010,Abriola2010}, and the spin for the isomeric state in  $^{76}$Cu is extracted from Ref.~\cite{deGrootePhD,deGroote2017}.}
\label{fig:level}
\end{figure*}

\section{Results}
\label{sec:results}
The results are summarised in Table~\ref{tab:results}. From the \linebreak ground- and isomeric-state masses, we determine the excitation energy for the isomer, $E_x=64.8(25)$~keV, for the first time. The new ground-state mass-excess value is around 30 keV lower than the literature value \cite{AME20}, which is mainly based on the most recent mass measurement at ISOLTRAP  \cite{Welker2017}. There the $^{76}$Cu ions were produced using neutron-induced fission on a uranium target and resonant laser ionization. The mass-excess value obtained at ISOLTRAP, $-50981.55(89)$~keV \cite{Welker2017}, is between the mass-excess values we determined for the two states at JYFLTRAP. This would suggest that a mixture of the states was measured at ISOLTRAP. The used quadrupolar excitation of 600~ms at ISOLTRAP \cite{pc} is not sufficient to resolve such close-lying states which explains why only one state was reported. We also note that the excitation energy matches with the $\gamma$-ray energies observed in the beta decay of $^{76}$Ni produced via in-flight fission of $^{238}$U at RIKEN \cite{Pedersen2019}. Three $\gamma$-ray peaks were identified as potential isomeric transitions in $^{76}$Cu, however, the 62~keV and 66~keV peaks were interpreted as a contamination from $^{75}$Cu \cite{Pedersen2019}.


To further elucidate the existence of two long-living states in $^{76}$Cu, additional measurements using the phase-imaging ion cyclotron resonance (PI-ICR) technique \cite{Eliseev2013, Eliseev2014,Nesterenko2018,Nesterenko2021} were performed. This technique is based onto the projection of the ion motion in the Penning trap on a position-sensitive microchannel-plate ion detector. The two states are clearly resolved (see Fig.~\ref{fig:pi-icr}). The PI-ICR measurements were performed using three different trap cycles with total lengths of 389, 489 and 1089~ms. During the trap cycle, the number of ions decreases according to the radioactive decay law and this allows for a rough half-life estimation of the states.  The number of ions collected with the 489-ms and 1089-ms cycles were normalised by taking into account the slightly different number of injections to the trap than in the 389-ms cycle (2479, 2855 and 1805 injections for the 389-ms, 489-ms and 1089-ms cycles, respectively). The fitted half-life for the isomer (see Fig.~\ref{fig:halflife}.a), $T_{1/2}$=672(110)ms, is in a very good agreement with the literature value, $T_{1/2}$=637.7(55)ms, \cite{Kondev2021}. 

The number of ground-state ions does not change much over time (see Fig.~\ref{fig:halflife}.a), but the trend between the 389-ms and 489-ms points is compatible with the literature half-life of 1.27(30)~s \cite{Winger1990}. It is likely that the isomeric state has an internal transition branch feeding the ground state, as suggested by the $\gamma$-ray peaks observed at 62~keV and 66~keV in Ref.~\cite{Pedersen2019} and the rather small spin difference between the two states (see the discussion later). The fitting of the ground-state half-life becomes impossible with three data points and three unknown parameters (the half-life, the initial number of ground-state ions and the internal transition branch of the isomer). Assuming literature half-lives for the two states, we obtain a range of possible solutions matching the observed ground-state data points (see Fig.~\ref{fig:halflife}.b). Based on the isomeric-state fit, the number of isomeric-state ions at time zero was around 1210(120). Thus, the isomer is much more strongly produced ($\gtrsim 90$~\%). The internal transition of the isomer has to be around $\approx 10-17$~\% to be compatible with the obtained results.  



\section{Discussion}
\label{sec:discussion}
Our measurements show that the dominantly produced shorter-living state is the isomer in $^{76}$Cu. The longer-living ground state has been missed in the previous experiments and rejected as non-existent in the NUBASE20 evaluation \cite{Kondev2021}. A recent CRIS study at ISOLDE measured only the shorter-living state and confirmed that it is a $3^-$ state \cite{deGrootePhD,deGroote2017}, however, it is plausible that some peaks belonging to the ground state have been missed in the regions not scanned intensively, in particular if the production rate of the ground state was much weaker. We note that the ISOLTRAP measurements employing the same production method at ISOLDE take a longer time, allowing the ground state to be populated via the internal transition branch. The information missed on the actual ground state of $^{76}$Cu can potentially impact on the interpretation of the trends in the copper isotopes \cite{Welker2017,deGroote2020}. The laser-spectroscopy studies have considered the $3^-$ state as the ground state, but the charge radius has been actually reported for the isomer. The conclusion on the observed unexpected reduction in the odd–even staggering of charge radii after $N=46$ \cite{deGroote2020} might thus change if the ground-state charge radius was significantly different from the isomer. 

It is somewhat surprising that the $3^-$ state at $64.8(25)$ ~keV is an isomer, which has a strong $\beta$-decay branch and does not immediately decay to the ground state with a presumed low-spin $J=(1-3)$. The spin assignment is based on the $\beta$-decay feeding to the first excited (2$^+$) state in $^{76}$Zn observed in Ref.~\cite{Winger1990}. However, isomerism with such a small spin difference is not unusual in the even-$A$ neutron-rich copper isotopes. For example, the $1^+$ isomer at 243-keV in $^{70}$Cu \cite{VanRoosbroeck2004,VanRoosbroeck2004prl} has a half-life of 6.6(2)~s and a $\beta$-decay branch of 93.2(9)\%, with the remaining fraction proceeding via a 141.3-keV $M2$ internal transition to the $3^-$ isomer. 

Weisskopf radiative transition probability estimates \cite{Weisskopf1951} would favor a $J=1^+$ assignment for the ground state of $^{76}$Cu. A 65-keV $M2$ transition in $^{76}$Cu would have a $\approx$1-ms half-life whereas $M1$ and $E1$ transitions are too fast with sub-ns half-lives. It should be noted that such a 65-keV $M2$ transition would be highly converted with a conversion coefficient which is defined as the ratio of the electron emission rate to the gamma emission rate $\alpha_{tot}=2.69(4)$ \cite{Kibedi2008}. If the ground-state spin was $1^+$, it would also allow the state to be populated directly via the beta decay of $^{76}$Ni. Low-spin states are often only weakly produced in direct fission \cite{Sears2021} but can be populated via beta decay, such as shown for the $(1^+$) ground state of $^{112}$Rh \cite{Hukkanen2023}. This could be an explanation why the weakly produced ground state has not been observed in many of the previous experiments. 

We have performed large-scale shell-model calculations and compared them to our experimental results to further understand the isomerism in $^{76}$Cu (see Fig.~\ref{fig:level}). The calculations with the LNPS interaction \cite{Lenzi2010} assume a $^{48}$Ca core, which consists of the negative-parity $pf$ shell for protons and $1p_{3/2}$, $1p_{1/2}$, $0f_{5/2}$, $0g_{9/2}$, and $1d_{5/2}$ orbits for neutrons. The $sdg$ neutron shell orbits missing from the LNPS calculations have been included in the PFSDG-U interaction \cite{Nowacki2016}. PFSDG-U calculations take $^{60}$Ca as an inert core, and block all cross-shell neutron and proton excitations. As can be seen in Fig.~\ref{fig:level}, both calculations predict almost degenerate $3^-$ and $4^-$ states and therefore, the ground state would be $4^-$ as the isomer has been verified as $3^-$ via laser spectroscopy \cite{deGroote2017,deGroote2020}. Both states have $\pi 1p_{1/2} \otimes (\nu 0g_{9/2})^{-3}$ as the leading configuration, with around 30~\% contribution to the whole wave function. 

 Experimental systematics of $^{72,74}$Cu would suggest $J=2^-$ for the ground state of $^{76}$Cu (see Fig.~\ref{fig:level}) while the Weiskopf estimates and low production would favour $J=1^+$. The shell-model calculations do not agree with the proposed spin assignments for the ground state of $^{76}$Cu. The first low-spin states in $^{76}$Cu are predicted at 960 keV (LNPS) and 763 keV (PFSDG-U), however, LNPS calculations predict a $1^+$ ground state for $^{72,74}$Cu. Previous shell-model calculations using JUN45 \cite{Honma2009} and jj44b (see the appendix of \cite{jj44b}) interactions have also had challenges to predict the ground-state spin-parity for $^{68,70,72,74}$Cu \cite{Vingerhoets2010}. Interestingly, the ground and isomeric states in the $N=47$ isotone $^{77}$Zn have spin-parities of $7/2^+$ and $1/2^-$ with leading $(\nu 0g_{9/2})^{-3}$ and $\nu 1p_{1/2}$ configurations, respectively \cite{Wraith2017}. Coupling a proton hole in $\pi 1p_{1/2}$ for $^{76}$Cu with these odd neutron configurations yields spin-parities of $3^-,4^-$ and $0^+,1^+$, compatible with the $3^-$ assignment of the isomer and tentative $(1^+)$ for the ground state in $^{76}$Cu. Further laser or decay spectroscopy studies are needed to make final conclusions on the spin-parity of the $^{76}$Cu ground state.


Although the absolute change in the mass-excess value of $^{76}$Cu is only 30 keV, it can have an impact on detailed calculations of nuclear structure as well as for astrophysical calculations. The impact of the $^{76}$Cu mass value from this work on core-collapse supernovae dynamics has been recently studied \cite{Giraud2022}. For the astrophysical rapid neutron capture ($r$) process \cite{Cowan2021}, sensitivity studies for the weak $r-$process scenario \cite{Surman2014} have shown that the neutron-capture reaction $^{76}$Cu$(n,\gamma)$ is one of the most influential reactions with an impact factor on the abundance of $F=$~25 when the rates were varied by a factor of 100. In addition, the influence of the presence of isomeric states on nucleosynthesis process in astrophysical environments is a fascinating and open question, and recent studies pointed out the importance of taking them into account in simulations \cite{Fujimoto2020,Misch2021}. The mass value impacts both the neutron-capture and its inverse reaction, photodissociation, reaction rates. The latter depends exponentially on the reaction $Q$-value: $\lambda_{\gamma,n} \propto$ $N_A<\sigma\nu>$~exp[$-Q/kT$]. The $Q$-value determined from this work, $Q$=5922.7(24)~keV is about 30-keV lower than the $Q$-value obtained using the $^{76}$Cu mass value published in AME2020 \cite{AME20}. At T=2~GK, the photodissociation rate calculated with the new $Q$-value is about 15\% higher than with the rate calculated using the AME2020 $Q$-value. 

In conclusion, we have verified the existence of two long-living states in $^{76}$Cu and determined the excitation energy of the isomer for the first time based on the measured masses of the ground and isomeric states. We have discovered that the ground state of $^{76}$Cu has been missed in all previous studies, leading to potentially erroneous conclusions on the evolution of ground-state properties of the copper isotopes. The key remaining uncertainty is related to the spin-parity assignment of the ground state of $^{76}$Cu and experimental effort should be addressed in this direction.

\section*{Acknowledgments}
This work has been supported by the Academy of Finland under the Finnish Centre of Excellence Program (Nuclear and Accelerator Based Physics Research at JYFL 2012-2017) and by the European Unions Horizon 2020 research and innovation programme grant agreement No 654002 (ENSAR2). A.K. acknowledges the support from the Academy of Finland under grant No. 275389 and D.N. and L.C. under grants No. 284516 and 312544. T.E. acknowledges the support from the Academy of Finland under grant No. 295207 and A.R. under grant No. 306980. A.K. and L.C. acknowledge the funding from the European Union’s Horizon 2020 research and innovation program under grant agreement No. 771036 (ERC CoG MAIDEN). We thank for the bilateral mobility grant from the Institut Fran$\text{\c{c}}$ais in Finland, the Embassy of France in Finland, the French Ministry of Higher Education and Research and the Finnish Society of Science and Letters. We are grateful for the mobility support from PICS MITICANS (Manipulation of Ions in Traps and Ion sourCes for Atomic and Nuclear Spectroscopy). S.G. thanks for the mobility grant from the EDPSIME.

\bibliography{biblio_76Cu.bib}

\end{document}